\documentclass[conference,10pt]{IEEEtran}
\IEEEoverridecommandlockouts

\pdfoutput=1
\usepackage{eqnarray}
\usepackage{cite}
\usepackage{amsmath,amssymb,amsfonts}
\usepackage{algorithmic}
\usepackage{graphicx}
\usepackage{textcomp}
\usepackage{xcolor}
\usepackage[utf8]{inputenc}
\usepackage{epsfig,graphics,amssymb} \usepackage{amscd} \usepackage{enumerate}
\usepackage{theorem} \usepackage{cite} \usepackage{stfloats} \usepackage{epsfig} \usepackage{verbatim}
\usepackage{graphicx} \usepackage{amssymb}  \usepackage{color}
\usepackage{bm}
\usepackage{algorithm}
\usepackage{CJK}

\begin{document}

\title{MmWave MIMO Communication with Semi-Passive RIS: A Low-Complexity Channel Estimation Scheme}

\author{\IEEEauthorblockN{Jiangfeng Hu\IEEEauthorrefmark{1}, Haifan Yin\IEEEauthorrefmark{1}, Emil Bj{\"o}rnson\IEEEauthorrefmark{2}\IEEEauthorrefmark{3}}

\IEEEauthorblockA{\IEEEauthorrefmark{1}Huazhong University of Science and Technology, Wuhan, China}
\IEEEauthorblockA{\IEEEauthorrefmark{2}KTH Royal Institute of Technology, Kista, Sweden}
\IEEEauthorblockA{\IEEEauthorrefmark{3}Link\"oping University, Link\"oping, Sweden}
Email: {\{jiangfenghu, yin\}@hust.edu.cn}, emilbjo@kth.se}

\maketitle

\begin{abstract}
Reconfigurable intelligent surfaces (RISs) have recently received widespread attention in the field of wireless communication. An RIS can be controlled to reflect incident waves from the transmitter towards the receiver; a feature that is believed to fundamentally contribute to beyond 5G wireless technology.
The typical RIS consists of entirely passive elements, which requires the high-dimensional channel estimation to be done elsewhere. Therefore, in this paper, we present a semi-passive large-scale RIS architecture equipped with only a small fraction of simplified receiver units with only 1-bit quantization. Based on this architecture, we first propose an alternating direction method of multipliers (ADMM)-based approach to recover the training signals at the passive RIS elements, We then obtain the global channel by combining a channel sparsification step with the generalized approximate message passing (GAMP) algorithm. Our proposed scheme exploits both the sparsity and low-rankness properties of the channel in the joint spatial-frequency domain of a wideband mmWave multiple-input-multiple-output (MIMO) communication system. Simulation results show that the proposed algorithm can significantly reduce the pilot signaling needed for accurate channel estimation and outperform previous methods, even with fewer receiver units.
\end{abstract}


\section{Introduction}
\par MmWave communication is one of the key components of 5G and beyond 5G cellular technology. In theory, mmWave bands may provide much higher throughput than conventional bands due to the large available bandwidth. However, this advantage is hard to utilize in practice due to two constraints: 1) the limited wireless coverage and 2) the limited power. The high penetration loss and lack of scattering paths make mmWave signals vulnerable to blockage. Even under ideal propagation conditions, the increased bandwidth leads to a reduction in signal-to-noise ratio (SNR) since it is hard to generate high-power mmWave signals.

A possible solution to these two problems is to deploy reconfigurable intelligent surfaces (RISs) \cite{Huang2018a,1di2020smart}, which can 1) circumvent blockage by intelligently reflecting wireless signals towards the desired receiver and 2) recycle existing radio waves to 
increase the SNR at the receiver without increasing the transmit power. A typical RIS consists of a large number of sub-wavelength-sized passive elements, whose impedance can be configured to control how incident waves are reflected.
While the RIS technology is in its infancy and the most promising use cases remain to be identified \cite{Bjornson2020a}, it is 
already regarded as a paradigm-shifting beyond 5G technology \cite{2wu2019towards}.
RIS might become essential to deal with the blockage/outage problem that frequently occurs in the mmWave communication system and ensure communication quality when the line-of-sight (LOS) link is unavailable.
This can be achieved with a limited energy consumption \cite{2wu2019towards,33pei2021ris}.

Despite its great potential, many practical problems remain to be solved. The key open problem is the acquisition of channel state information (CSI) \cite{Bjornson2020a}, which is the basis for the RIS configuration for intelligent reflection.
The channel estimation problem is challenging (even when channel reciprocity exists \cite{33pei2021ris}) since once the RIS is involved in the communication system, its large number of reflecting elements will introduce an additional high-dimensional cascaded channel. The authors of \cite{6liu2020matrix} and \cite{7he2019cascaded} adopted matrix factorization/completion methods to decompose the  cascaded channel among base station (BS)-RIS-user equipment (UE), regarding the distribution of the channel as prior information. The authors of \cite{8hu2019two,9wan2020broadband} assumed that the channel between BS and RIS is quasi-static, requiring infrequent estimation, while the relatively fast-changing RIS-UE channel should be estimated more frequently. However, due to the passive nature of the RIS elements, the two parts of the cascaded channel cannot be decoupled very well. This will not only lead to a scaling ambiguity but an immense computational burden on the transceivers. 

Given this dilemma, a few researchers have recently proposed to replace a small fraction of the RIS elements with receive antennas and let the RIS undertake some computing tasks. Thus, the RIS can adjust its reflection coefficients according to the CSI it estimates in a real-time manner instead of being controlled by other transceivers. This kind of RIS configuration was originally proposed in \cite{11taha2019enabling} (termed semi-passive RIS in \cite{10wu2020intelligent}), by which we can well decouple the aforementioned cascaded channel. Early explorations \cite{11taha2019enabling,34liu2020deep,13alexandropoulos2020hardware,35zhang2020deep}, which mainly assumed the single-input-single-output (SISO) scenario, have demonstrated that the global CSI can be recovered by activating a small proportion of RIS units combined with deep learning (DL), compressed sensing (CS), and other signal processing techniques. Still, they face the problems of either obtaining labeled training data or high computational complexity.

Inspired by the recently proposed random spatial sampling techniques \cite{14vlachos2019wideband}, in this paper, we present a novel architecture of a large-scale RIS. To keep the hardware cost and energy consumption low, we consider having very few simplified receiver units at the RIS with only 1-bit quantization, which are randomly connected to a small fraction of RIS elements during the channel estimation stage. Capitalizing on the mmWave channel properties and the proposed architecture, we present an efficient ADMM-based  quantified matrix completion (MC) approach, which exploits the low-rankness property of the channel \cite{18yin2013coordinated} when recovering the pilot signals received by passive RIS elements. To reconstruct the channel with the recovered training signals, we propose to combine the GAMP algorithm with a channel sparsification step that exploits the channel sparsity in three domains: the transmit angular domain, the receive angular domain, and the frequency domain. The proposed scheme operates in a broadband mmWave MIMO communication system, which is more practical but rarely studied in current RIS literature.

\emph{Notations:} We use bold-face to denote vectors and matrices. For a matrix ${\bf{X}}$, ${\rVert{\bf{X}\|_*}}$, ${\rVert{\bf{X}\|_\infty}}$, ${({\mathbf{X}})^T}$, ${({\mathbf{\bar X}})}$, and${({\mathbf{X}})^*}$, denote the nuclear norm, infinity norm, transpose, conjugate, and conjugate transpose of it respectively. ${\mathbb{E}\{{\cdot}\}}$ is the expectation. The Kronecker and Hadamard product are denoted as $ \otimes $ and $\odot$, respectively. Vectorization of matrix ${\bf{X}}$ is represented by $\text{vec}({\bf{X}})$, and $\text{unvec}( \cdot )$ is the inverse operation of $\text{vec}( \cdot )$.

\section{System Model}

\par
\begin{figure}[h]
  \centering
  \includegraphics[width=3.2in]{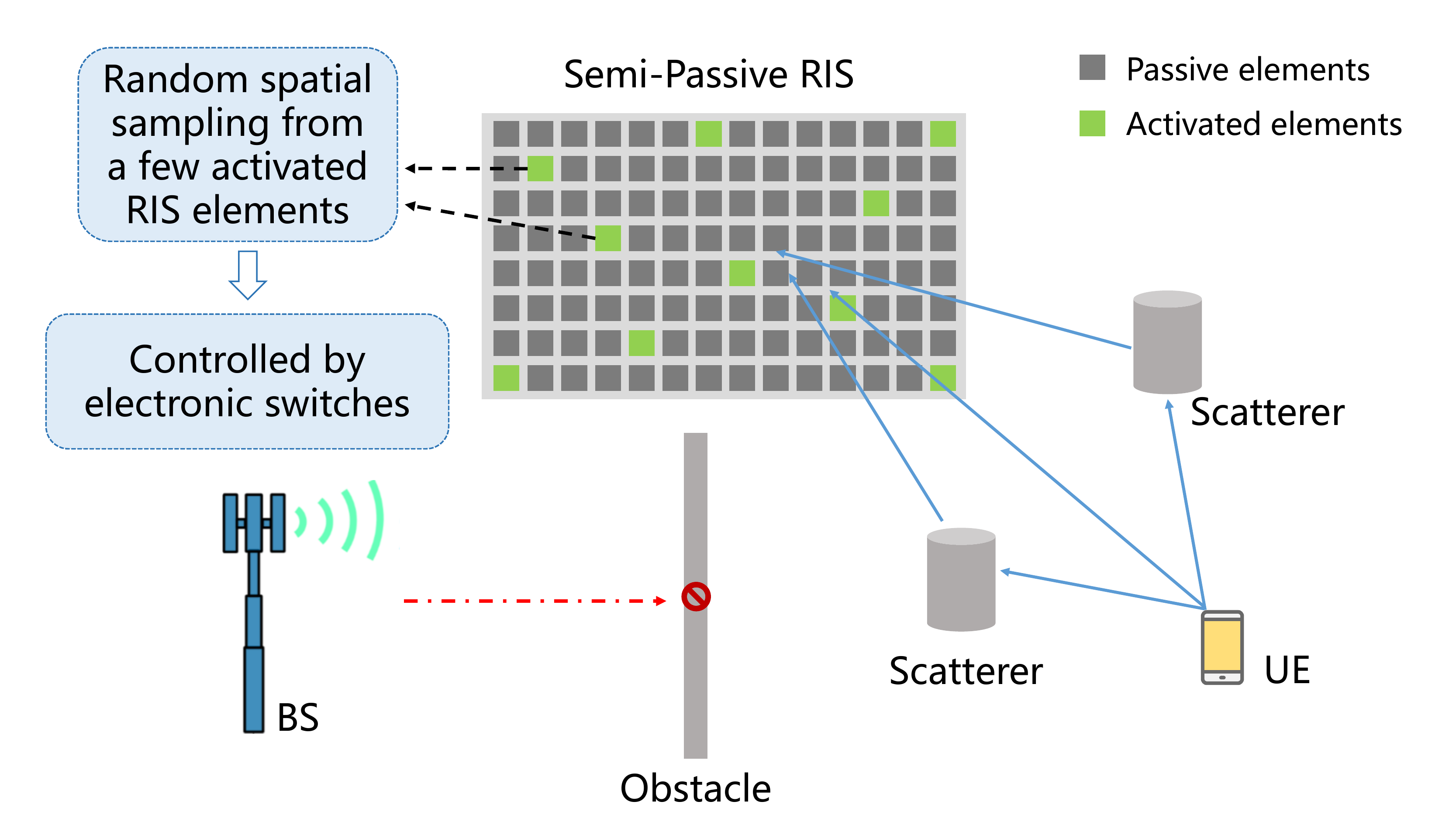}\\
  \caption{Schematic diagram of proposed semi-passive-RIS-aided communication system, where a small fraction of simplified receiver units will be randomly connected to RIS units to receive impinging signals during the channel estimation phase.} \label{fig:msePartialOverlapping}
\end{figure}

 We consider a large-scale semi-passive-RIS-aided broadband communication system operating in the mmWave band, where the LOS link is blocked due to blockage, as illustrated in Fig.~\ref{fig:msePartialOverlapping}. Compared with RISs operating in sub-6 GHz, the total number of subpaths in the  mmWave channel is relatively small due to the limited scattering environment. 
To keep the hardware cost and energy consumption of the semi-passive RIS low, there are only a few simplified receiver units with 1-bit quantization on the RIS side, which can be mainly realized by mixers and low-precision phase/amplitude detectors with low cost. An alternative scheme of the receiver units maybe only radio frequency sampling ADCs. These are randomly connected to a fraction of the RIS elements by electronic switches that change during the channel estimation phase. The system operates in time-division duplex (TDD) mode, and the uplink (UL) and downlink (DL) occupy the same bandwidth, consisting of $N_k$ subcarriers with spacing $\Delta f$. The BS, RIS, and UE are all equipped with uniform planar arrays (UPAs). 

The BS-RIS and RIS-UE channels are modeled using geometric far-field models from \cite{17yin2020addressing}. Since both of the two parts have similar characteristics and the locations of BS and RIS are usually fixed, thus the channel between them is relatively easier to estimate. Therefore, we mainly focus on the channel estimation problem about the UL between RIS and UE. The element number of transmitter (UE) and receiver (RIS) are ${N_t}$ and ${N_r}$, respectively. The channel is composed of $N_{cl}$ scattering clusters, each consisting of $N_{sp}$ subpaths. For the $(m,n)$-th subpath, $\alpha_{n, m}$, $\tau_{n, m}$, $\varphi_{n, m}^{r}\left(\theta_{n, m}^{r}\right)$, $\varphi_{n, m}^{t}\left(\theta_{n, m}^{t}\right)$ denote the complex gain, the delay, the azimuth (or zenith) angle of arrival, and the azimuth (or zenith) angle of departure, respectively. We consider a UPA with isotropic antennas located on YZ plane. Its array response vector is given by 

\begin{equation}
    \centering
    \label{equ1}
    \mathbf{a}(\theta, \phi)=\mathbf{a}_{h}(\theta, \phi) \otimes \mathbf{a}_{v}(\theta),
\end{equation}
where $\varphi$ is the azimuth angle, $\theta$ is the zenith angle, and
\begin{equation}
    \centering
    \label{equ2}
{{\bf{a}}_h}(\theta ,\phi) = {[1\;{e^{j2\pi \frac{{{D_h}\sin (\theta)\sin (\phi )}}{{{\lambda _0}}}}}\;\!\!\!\; \cdots \;\;\!\!\!{e^{j2\pi \frac{{({N_h} - 1){D_h}\sin (\theta )\sin (\phi)}}{{{\lambda _0}}}}}]^T}
\end{equation}
\begin{equation}
    \centering
    \label{equ3}
{{\bf{a}}_v}(\theta ) = {[1\;\;{e^{j2\pi \frac{{{D_v}\cos (\theta )}}{{{\lambda _0}}}}}\;\; \cdots \;\;{e^{j2\pi \frac{{({N_v} - 1){D_v}\cos (\theta )}}{{{\lambda _0}}}}}]^T},
\end{equation}and ${\lambda _0},\;{D_h} ({D_v}),\;{N_h} ({N_v})$ denote the wavelength, horizontal (vertical) antenna spacing and number of antenna, respectively. The channel transfer function on the $k$-th subcarrier is represented as:
\begin{equation}
    \centering
    \label{equ4}
\begin{array}{l}
{\bf{H}}[{f_k}] = \sum\limits_{n = 1}^{{N_{cl}}} {\sum\limits_{m = 1}^{{N_{sp}}} {{\alpha _{n,m}}} } {{\bf{a}}_r}(\varphi _{n,m}^r,\theta _{n,m}^r){\bf{a}}_t^*(\varphi _{n,m}^t,\theta _{n,m}^t)\\
\\
\;\;\;\;\;\;\;\;\;\;\;\;\; \times {e^{ - j2\pi {f_k}{\tau _{m,n}}}},\;\;\;0 \le k \le {N_k},
\end{array},
\end{equation}
thus its rank is limited by
\begin{align}
\centering
\label{equ5}
& \text{rank}({\bf{H}}[{f_k}])  \\
&= \text{rank}(\sum\limits_{n = 1}^{{N_{cl}}} {\sum\limits_{m = 1}^{{N_{sp}}} {{{\bf{a}}_r}(\varphi _{n,m}^r,\theta _{n,m}^r){\bf{a}}_t^*(\varphi _{n,m}^t,\theta _{n,m}^t))} } \\
&\le {N_{cl}} \cdot {N_{sp}} = {N_{path}},
\end{align} and the total channel at all $N_k$ subcarriers is given by
\begin{equation}
    \centering
    \label{equ6}
{\bf{H}} = {\left[ {{{({\bf{H}}[{f_1}])}^T}\;{{({\bf{H}}[{f_2}])}^T}\; \cdots \;{{({\bf{H}}[{f_{{N_k}}}])}^T}} \right]^T} \in {\mathbb{C}^{{N_k}{N_r} \times {N_t} }},
\end{equation}
which is also low-rank due to (7) and the large available bandwidth of mmWave.

\subsection {Channel sparsification method }
\par We will exploit the mmWave channel sparsity inherent in the transmit/receive angular domains and the frequency domain during the channel estimation phase. To this end, we first transform the channel transfer function to the angular domain using discrete Fourier transform (DFT) matrix: 
\begin{equation}
\centering
\label{equ7}
\mathbf{A}\left[f_{k}\right]=\mathbf{B}_{N_{r}}^{*} \mathbf{H}\left[f_{k}\right] \mathbf{B}_{N_{t}}
,\end{equation}
where $\mathbf{B}_{N_{r}}=\mathbf{D}_{N_{r}}^{v} \otimes \mathbf{D}_{N_{r}}^{h}$, $\mathbf{B}_{N_{t}}=\mathbf{D}_{N_{t}}^{v} \otimes \mathbf{D}_{N_{t}}^{h}$, and $\mathbf{D}_{N} \in \mathbb{C}^{N \times N}$ is a DFT matrix of $N$ dimensions, which reflects the geometric characteristics of the antenna array, and can be also regarded as an orthogonal projection basis \cite{17yin2020addressing}. Thus, after projection, only a few elements in $\mathbf{A}\left[f_{k}\right]$ have relatively high magnitude. Note that the transformation matrix itself will affect the sparsity and energy leakage of $\mathbf{A}\left[f_{k}\right]$. Still, considering the fast algorithm implemented by fast Fourier transform (FFT), we only adopt the standard DFT matrix in this paper. Moreover, in order to simultaneously exploit the broadband mmWave channel sparsity in terms of multipath delays, we further transform  $\mathbf{A}\left[f_{k}\right]$ to the 
delay domain as follows:
\begin{equation}
\centering
\label{equ8}
\tilde {\mathbf{A}}=\mathbf{D}_{N_{k}} \mathbf{A}
,\end{equation}
where ${\bf{A}} = {\left[ {\text{vec}({\bf{A}}[{f_1}])\;\text{vec}({\bf{A}}[{f_2}])\; \cdots \text{vec}({\bf{A}}[{f_{{N_k}}}])} \right]^T}$. Finally, the joint spatial-frequency orthogonal basis can be formulated as 
\begin{equation}
\centering
\label{equ9}
\mathbf{S}=\mathbf{D}_{N_{k}} \otimes \mathbf{B}_{N_{r}} \otimes \mathbf{B}_{N_{t}}^{*}
,\end{equation}
and the channel sparsification is done by projecting the wideband mmWave channel onto it as ${\bf{x}} = {\bf{S}}\text{vec}({\bf{H}})$. Thus, the channel coefficients obtained after that are much sparser than before \cite{18yin2013coordinated}. Note that, in practice, $N_k$ can also be the number of resource blocks (RBs) or groups of consecutive RBs depending on the Sounding Reference Signal (SRS) frequency structure, which will lead to a much smaller value than the total number of subcarriers. 

\subsection {Fast Implementation}

Considering a block fading channel whose coefficients remain unchanged for a block of consecutive symbols, here we take the channel between the RIS and the UE as an example. In the UL, a training block consisting of $N_p$ pilots is denoted as $\mathbf{T}$, and the signals that the RIS can receive within this block are written as 
\begin{equation}
\centering
\label{equ10}
{\bf{Z}} = ({{\bf{B}}_{N_r}} \otimes {{\bf{D}}_{{N_k}}}){\bf{XC}}
,\end{equation}
where $ {\bf{C}}={\bf{\bar B}}_{{N_t}} \mathbf{T}$, and
\begin{equation}
\centering  
\label{equ11}
{\bf{X}} = {\left[ {{{({\bf{X}}[{f_1}])}^T}\;{{({\bf{X}}[{f_2}])}^T}\; \cdots \;{{({\bf{X}}[{f_{{N_k}}}])}^T}} \right]^T},
\end{equation} which is attained by $\mathbf{X}=\text {unvec}(\mathbf{x}) \in \mathbb{C}^{N_r N_k \times N_t}$. We adopt the widely used Zadoff-Chu (ZC) sequence as pilot signal, which has an autocorrelation property and low peak power ratio. We rewrite $\mathbf{X}$ and $\mathbf{Z}$ in vector forms as follows:
\begin{equation}
\centering
\label{equ12}
\begin{aligned}
{\bf{z}} &= {\rm{vec}}({\bf{Z}})\\
 &= ({{\bf{C}}^T} \otimes ({{\bf{B}}_{{N_r}}} \otimes {{\bf{D}}_{{N_k}}})){\rm{vec}}({\bf{X}})\\
 &= {\boldsymbol{\psi}{\bf{x}}}
\end{aligned},
\end{equation}
where $\boldsymbol{\psi}$ is the dictionary matrix in the compressed sensing problem. Thus, we have transformed the channel estimation problem into a sparse vector recovery problem. Considering the massive number of elements in the RIS, the matrix $\boldsymbol{\psi}$ will become too large to store in memory and perform subsequent calculations. To cope with this problem, we represent $\boldsymbol{\psi}$ as an implicit operator in subsequent simulations, of which the components can be effectively calculated by FFT algorithms. Note that due to the UPA we consider, where $\mathbf{B}_{N}=\mathbf{D}_{N}^{v} \otimes \mathbf{D}_{N}^{h} $. We can still adopt the FFT by exploiting the property of the Kronecker product. To put it in practical forms, for an arbitrary vector $\mathbf{v}$, we have
\begin{equation}
\centering
\label{equ13}
\mathbf{B}_{N} \mathbf{v}=\left(\mathbf{D}_{N}^{v} \otimes \mathbf{D}_{N}^{h}\right) \mathbf{v}=\operatorname{vec}\left(\mathbf{D}_{N}^{h} \mathbf{V} \mathbf{D}_{N}^{v}\right)
,\end{equation}
where $\mathbf{V}=\text {unvec}(\mathbf{v}) \in \mathbb{C}^{N \times N}$. In this way, a column-wise two-dimensional FFT can be performed as well.

\section{Proposed channel estimation algorithm}
\par Based on the system model derived in the previous section, we propose an effective  scheme for semi-passive-RIS-aided channel estimation, which exploits the joint spatial-frequency sparsity and low-rankness property of the mmWave channel, combined with the spatial random sampling technique and nonlinear signal reconstruction algorithm.

Recall that there are only a few simplified receiver units with 1-bit quantization in the RIS, which will make the channel estimation problem interesting and challenging. Nevertheless, the proposed scheme turns out to be effective. To ensure the accuracy of the signal reconstruction, we carry out random sampling within each training signal block. The corresponding physical realization is to randomly control a proportion of the RIS elements and change that portion over the channel training time by electronic switches. Considering the effects of random sampling and noise (both the thermal and quantization noise), the total received signal in the RIS is
\begin{equation}
\centering
\label{equ14}
{\bf{Y}_\Omega}=\boldsymbol{\Omega} \odot Q(\mathbf{Z}+\mathbf{W}),
\end{equation}
where $\boldsymbol{\Omega}$ is the sampling matrix composed of $\{0,1\}$ entries, ${\bf{W}}$ represents the additive thermal noise, and $Q( \cdot )$ denotes the quantization function. The matrix ${{\bf{Y}_\Omega} }$ is incomplete due to the low spatial sampling rate we take. Due to the mmWave channel sparsity, the rank of the received signal matrix is also limited:
\begin{equation}
\centering
\label{equ15}
\text{rank}({\bf{Z}}) \le \text{rank}({\bf{X}}) \le {N_{path}}
.\end{equation} 
Therefore, if the impact of quantization effects is not considered for the time being, the traditional MC methods such as \cite{20cai2010singular,21lin2009augmented}, which can recover a low-rank matrix from an incomplete set of measurements, are suitable to recover the entire signal matrix from a small fraction of observations. Another reason we choose to combine the proposed semi-passive RIS hardware with MC methods is that, with the iterative thresholding step in  \cite{20cai2010singular}, we are able to recover an $n \times n$ matrix with rank $r$ from $\mathcal{O}(n r \rm{poly}\log (n))$ uniformly random samples. Therefore, once the rank of the matrix to be recovered is fixed, the larger the dimension of it, the lower proportion of the data needs to be sampled. 
Hence, in a given setup, it might be possible to both increase the number of RIS elements, to capture more signal energy, and simultaneously keep the rank constant so that the number of receiver units can remain the same. 

However, when it comes to our case of recovering a 1-bit quantized received signal matrix, the MC methods above cannot be applied due to their strict constraints (i.e., the incoherence) on the original matrix and their sensitivity to noise (especially quantization noise). Hence, we choose to relax the 
quantified MC problem in a way inspired by \cite{36ni2016optimal}:
\begin{equation}
\centering
\label{equ16}
\mathop {\min }\limits_{{\bf{\hat Y}}} \rVert{ {{\bf{\hat Y}} - {{\bf{Y}}_\Omega }}\|_F^2}, \ \rm{s.t.} {  \rVert{\bf{\hat Y}}\|_*} \le \sigma , {  \rVert{\bf{\hat Y}}\|_\infty} \le \gamma, \end{equation} which can be cast into the following separable form:
\begin{equation}
\centering
\label{equ17}
\mathop {\min }\limits_{{\bf{\hat Y}}} \rVert{ {{\bf{\hat Y}} - {{\bf{Y}}_\Omega }}\|_F^2}, \ \rm{s.t.} {  \rVert{\bf{\hat Y}}\|_*} \le \sigma , {  \rVert{\bf{\bar Y}}\|_\infty} \le \gamma, {\bf{\hat Y}} = {\bf{\bar Y}}. \end{equation}
Then, an ADMM-based approach can be applied, with the corresponding augmented Lagrangian function given by
\begin{equation}
\centering
\label{equ18}
\begin{aligned}
{L_\mu }({\bf{F}},{\bf{\hat Y}},{\bf{\bar Y}}) &=
\rVert{ {{\bf{\hat Y}} - {{\bf{Y}}_\Omega}}\|_F^2} - \langle {{\bf{F}},{\bf{\hat Y}} - {\bf{\bar Y}}} \rangle \\ & + \frac{\mu }{2} \rVert{ {{\bf{\hat Y}} - {{\bf{\bar Y}} }}\|_F^2} + \alpha {  \rVert{\bf{\hat Y}}\|_*}  + \beta {{\bf{1}}_{[{{\rVert{\bf{\bar Y}}\|_\infty} } \le \gamma ]} },
\end{aligned}
\end{equation}
where the indicator function is defined as
\begin{equation}
\centering
\label{equ19}
{{\bf{1}}_{[{{\rVert{\bf{\bar Y}}\|_\infty} } \le \gamma ]}} = \left\{ \begin{array}{l}
0,\ {\rVert{\bf{\bar Y}}\|_\infty } \le \gamma,  \\
1,\ {\rm{ otherwise}},
\end{array} \right.
\end{equation}
and the element-wise infinity norm constraint is added to avoid the overly spiky matrix in the iterative process, which is a much-relaxed constraint compared to the incoherence condition. Therefore, the solution can be found by alternately minimize the Lagrangian function above, and the detailed steps are described in {{Algorithm 1}}. Note that different from the traditional parameters updating order, here we adopt the strategy in \cite{66cai2013proximal} to accelerate the convergence rate.

The problem to be solved is, under the condition that the recovered received signal vector $\hat{\mathbf{y}}=\operatorname{vec}(\hat{\mathbf{Y}})$, and the mixing dictionary matrix $\boldsymbol{\psi}$ are known, how to accurately estimate the vectorized sparse channel $\mathbf{x}$, which can be regarded as a noisy quantized CS problem. We adopt here the GAMP algorithm, which transforms the vector estimation problem into a series of constant estimation problems. It has the advantages of lightweight calculation and fast convergence, and is very suitable for the reconstruction of component-wise nonlinear signals \cite{15mo2017channel}. However, the preset channel sparsity rate $\eta$ will have a substantial impact on the results of the methods above, which is defined as follows: 
For a K-sparse vector ${\bf{x}}$, i.e., ${{\bf{x}} \in {\mathbb{C}^N}}$ with K nonzero entries, then its sparsity rate is defined as $\eta  = K/N$.

In practice, the sparsity rate of the channel vector is a priori unknown. Therefore, we chose to exploit the Expectation-Maximization (EM) method to update the relevant parameters as in \cite{19vila2013expectation}, including the channel sparsity rate at each step of the M-step iteration. More specifically, we propose the following channel estimation algorithm combined with the aforementioned semi-passive RIS hardware in this paper.

Since GAMP belongs to the Bayes-like estimation framework, we need to determine the corresponding marginal posterior distributions by assuming a prior one first. To achieve a good fit to the true distribution of the vectorized channel coefficients, here we assume that each element of it satisfies the following Gaussian-Mixture distribution: 
\begin{equation}
\centering
\label{equ20}
{p_X}(x;\eta ,\omega ,\theta ,\phi ) = (1 - \eta )\delta (x) + \eta \sum\limits_{l = 1}^L {{\omega _l}{{\cal N}}} (x;{\theta _l},{\phi _l}),
\end{equation}
where $\delta ( \cdot )$ is the Dirac delta function and ${\cal N}(x;a,b)$ is the probability density function of a Gaussian random variable $x$ with mean $a$ and variance $b$. We denote by $\omega_{l}, \theta_{l}, \phi_{l}$ as the weight, mean, variance for the $l$-th GM component, respectively. Since the noise is independent of the signal, it is assumed to be i.i.d. Gaussian, and satisfies ${\cal N}(w;0,\varphi )$. The above parameters are known a priori in the original GAMP algorithm but need to be estimated iteratively in EM-GAMP. For the ease of presentation, we denote them together as $\mathbf{q} = [\eta, \omega, \theta, \phi, \varphi]$. Since the true marginal posterior $p\left(z_{m} \mid \mathbf{y} ; \mathbf{q}\right)$ is usually analytically intractable and computationally prohibitive, the GAMP-based algorithm approximates it by 
\begin{equation}
\label{equ21}
\begin{array}{l}

{p_{Z|{\bf{\hat Y}}}}({z_m}|{\bf{\hat y}};{{\hat p}_m},\mu _m^p,{\bf{q}})
 \buildrel \Delta \over = \frac{{{p_{\hat Y|Z}}({{\hat y}_m}|{z_m};{\bf{q}}){\cal N}({z_m};{{\hat p}_m},\mu _m^p)}}{{\int\limits_z {{p_{\hat Y|Z}}({{\hat y}_m}|z;{\bf{q}})} {\cal N}(z;{{\hat p}_m},\mu _m^p)}},

\end{array}
\end{equation} 
 and then evaluate the means and variances of it, where   $z_{m}={\bf{a}}_m^T \bf{x}$ in (14), i.e., ${\bf{a}}_m^T$  is the $m^{t h}$  row of dictionary matrix $\boldsymbol{\psi}$. $\hat{p}_{m}$ and $\mu_{m}^{p}$ are the mean and variance of $z_{m}$, which will change during iterations. However, since our system output $\hat{\mathbf{y}}$ (received  signal we obtain from MC) is 1-bit quantized, it is different from the additive white Gaussian noise (AWGN) output assumption in \cite{25rangan2011generalized}, here we have 
\begin{equation}
\begin{aligned}
\centering
\label{equ22}
{p_{\hat{Y} \mid Z\left(\hat{y}_{m} \mid z_{m}\right)} = {P}\left\{\hat{y}_{m} \mid z_{m}\right\}={P}\left\{Q\left(z_{m}+w_{m}\right) \mid z_{m}\right\}}.
\end{aligned}
\end{equation}

Similarly,  the GAMP algorithm approximates the true marginal posterior  $p\left(x_{n}\mid \mathbf{y} ; \mathbf{q}\right)$  by 
\begin{equation}
\centering
\label{euq22}
{p_{X|{\bf{\hat Y}}}}({x_n}|{\bf{\hat y}};{\hat r_n},\mu _n^r,{\bf{q}}) \buildrel \Delta \over = \frac{{{p_X}({x_n};{\bf{q}}){\cal N}({x_n};{{\hat r}_n},\mu _n^r)}}{{\int\limits_x {{p_X}({x_n};{\bf{q}})} {\cal N}(x;{{\hat r}_n},\mu _n^r)}}
,\end{equation} where  $\hat{r}_{n}$ and $\mu_{n}^{r}$  are the mean and variance of $x_{n}$. By combining (25) with (22), then we have the detailed GM-GAMP approximated posterior as follows:
\begin{equation}
\label{euq23}
\begin{array}{l}
{p_{X|\bf{\hat{Y}}}}({x_n}|{\bf\hat{{y}}};{{\hat r}_n},\mu _n^r,{\bf{q}})\\
 = \left( {(1 - \lambda )\delta ({x_n}) + \lambda \sum\limits_{l = 1}^L {{w_l}{\cal N}({x_n};\theta ,{\phi _l})} } \right)\frac{{{\cal N}({x_n};{{\hat r}_n},\mu _n^r)}}{{{\zeta _n}}}\\
 = (1 - {\pi _n})\delta ({x_n}) + {\pi _n}\sum\limits_{l = 1}^L {{{\bar \beta }_{n,l}}{\cal N}({x_n};{\gamma _{n,l}},{\nu _{n,l}})},
\end{array}
\end{equation} where $\zeta_n$ is a normalization factor, and the parameters ${{{\bar\beta _{n,l}}}}, {{\gamma _{n,l}}}, {{\nu _{n,l}}},  {\pi _n}$ are all 
dependent on ${{\hat r}_n},\mu _n^r,{\bf{q}}$ \cite{25rangan2011generalized}. With (26) obtained, each ${x_n}$ in ${\bf{x}}$ can be easily estimated by
\begin{equation}
\centering
\label{euq24}
{\hat{x}_n }={{\mathbb{E}}_{X|{\bf{\hat Y}}}}({x_n}|{\bf{\hat y}};{\hat r_n},\mu _n^r,{\bf{q}})={\hat r_n} + \frac{{\mu _n^r}}{{\mu _n^r + \varphi }}({y_n} - {\hat r_n}).
\end{equation}

In every iteration of the GAMP algorithm, we exploit the EM method to update the parameters in  $\mathbf{q}$ as well as $\hat{\mathbf{x}}$, and repeat until convergence. Note that all of the quantities needed for the EM updates are already computed by the GAMP algorithm, making the overall process computationally efficient \cite{19vila2013expectation}. Owing to the aforementioned fast implementation method, the total computational complexity of EM-GAMP is reduced to $\mathcal{O}({N_r}{N_t}{N_k}\log ({N_r}{N_t}{N_k}))$. The main complexity of the MC process is due to the partial singular value decomposition, whose complexity is $\mathcal{O}(n{r^2})$ for an $m \times n$ matrix $(m > n)$ with rank $r$. 
To ensure the possibility of accurate recovery, we choose to reshape the received signal matrix to a dimension of ${\mathbb{C}^{{N_r} \times {N_p} {N_k}}}$. 

The proposed algorithm is summarized in {{Algorithm 1}}:
\columnsep 0.2in
\begin{algorithm}[h]
\caption{Proposed channel estimation algorithm for semi-passive-RIS-aided broadband mmWave MIMO system}

\begin{algorithmic}[1]
\renewcommand{\algorithmicrequire}{\textbf{Input:}}
\renewcommand{\algorithmicensure}{\textbf{Output:}}
\REQUIRE the random measurement matrix ${{\bf{Y}}_\Omega }$, pilot block ${\bf{T}}$, DFT matries with required dimensions, the threshold $\varepsilon_1, \varepsilon_2$ and iteration numbers $k_{\max},t_{\max}$ for stop criterion.
\ENSURE reconsructed channel vector ${\bf{\hat x}}. $
\label{alg:1}

\STATE{\textbf{Initialization}: set ${{\bf{\hat q}}_0}$, $\alpha $, $\beta $, ${\mu _0}$, $\sigma $, $\gamma $, ${{\bf{F}}^0}$ properly, and select ${  \rVert{\bf{\hat Y}}\|_*} \le \sigma $, ${  \rVert{\bf{\bar Y}}\|_\infty} \le \gamma $. set ${\bf{\hat x}}$ to a random vector.
}

\STATE {\textbf{for} $k = 1:k_{\max}$ }
\STATE \quad{\bf Minimization of ${L_\mu }({\bf{F}},{\bf{\hat Y}}, {\bf{\bar Y}})$  over ${\bf{\bar Y}}$:}
\STATE \hspace{-2.5mm} \quad
\begin{small}{$\begin{array}{l}
{{{\bf{\bar Y}}}^{k + 1}} = 
\mathop {\min }\limits_{{\bf{\tilde Y}}} \rVert{{{\bf{\bar Y}} - (2{{\bf{Y}}_\Omega } + 2{{\bf{F}}^k} + (2 + {\mu _k}){{{\bf{\hat Y}}}^k})/(4 + {\mu _k})}\|_F^2{\rm{ }}}\\\end{array}$ } \\\quad \! s.t.${{  \rVert{\bf{\bar Y}}\|_\infty} \le \gamma  }; $\end{small}

\STATE \quad{\bf{Update the  dual variable:}}
\STATE \quad${{\bf{F}}^{k + 1}} = {{\bf{F}}^k} + {\mu _k}({\bf{\hat Y}} - {{\bf{\bar Y}}^{k + 1}})$;
\STATE \quad{\bf Minimization of ${L_\mu }({\bf{F}},{\bf{\hat Y}},{\bf{\bar Y}})$ over ${\bf{\hat Y}}$:}
\STATE \quad \begin{small}{${{{\bf{\hat Y}}}^{k + 1}} = \mathop {\min }\limits_{{\bf{\hat Y}}} \rVert{ {{\bf{\hat Y}} - {{{\bf{\bar Y}}}^{k + 1}} + 2{{\bf{F}}^{k + 1}}/(2 + {\mu _k})}\|_F^2}$ \\ \! \quad s.t.${  \rVert{\bf{\hat Y}}\|_*} \le \sigma $}; \end{small}

\STATE \quad{${\mu _{k + 1}} = 1.01{\mu _k}$};
\STATE \quad{${k = k + 1}$};
\STATE \quad{\bf{if}} ${\rVert{{{{{\bf{\hat Y}}}^k} - {{\bf{Y}}_\Omega }}\|_F}} \le {\varepsilon _1}$ {\bf{{or} ${\rVert{ {{{{\bf{\hat Y}}}^k} - {{{\bf{\bar Y}}}^k}}}\|_F} \le {\varepsilon _1}$}, break;}
\STATE {\textbf{end for} }
\STATE {${\bf{\hat y}} = \text{vec}({\bf{\hat Y}})$};

\STATE {\bf{for} $t = 1:t_{\max}$ }

\STATE \quad {{\bf{E-step}}: Compute ${{\bf{\hat x}}_t}$  from ${p_{X|{\bf{\hat Y}}}}({x_n}|{\bf{\hat y}};{\hat r_n},\mu _n^r,{\bf{q}})$ by \\ \quad exploiting GAMP algorithm; }

\STATE \quad {{\bf{M-step}}: update ${{\bf{\hat q}}_t}$ as described in \cite{19vila2013expectation};}
\STATE \quad{t = t+1; }
\STATE \quad{\bf{if $\left\| {{{{\bf{\hat x}}}_t} - {{{\bf{\hat x}}}_{t - 1}}} \right\| \le {\varepsilon _2}$, break;}}
\STATE {\textbf{end for} }
\STATE {\textbf{return} ${\bf{\hat x}} = {{\bf{\hat x}}_t}$};
\end{algorithmic}
\end{algorithm}

\section{Simulation Results}
We consider a UL UE-RIS broadband mmWave channel estimation problem in a single cell. The system operates in 28 GHz band, with a total bandwidth of about 200 MHz. Orthogonal Frequency Division Multiplexing (OFDM) modulation is assumed, as in practical system and the subcarrier spacing is 60 kHz. The bandwidth is divided into 64 sub-bands, with each sub-band composed of 4 consecutive RBs. The pilot density in the frequency domain is one for each sub-band. The channel has 4 clusters, each consisting of 5 paths and 7.5 degrees of azimuth and elevation angular spread. The transmitting UE is equipped with a normal $2 \times 8$ UPA, and the receiver (semi-passive RIS) is equipped with a $32\times32$ UPA and a small proportion (8\%) of simplified receiver units with 1-bit quantization. We compare the performance of the proposed joint channel estimation scheme with several traditional algorithms that assume all of the RIS elements can receive signals under exhaustive Monte-Carlo simulations. The performance metric is the normalized mean square error (MSE), which is given by ${\bf{{\mathbb E}}}\{ {\| {{\bf{\tilde x}} - {\bf{x}}} \|^2/\| {\bf{x}} \|^2} \}$, where $\tilde{\mathbf{x}}$ is the normalized estimation of $\mathbf{x}$. And the SNR is defined as ${\rm{SNR}} \buildrel \Delta \over = {\mathbb{E}}\{ {{{\| {\bf{z}} \|}^2}} \}{\mathbb{/E}}\{ {{{\| {\bf{w}} \|}^2}} \}$, where $\mathbf{w}=\text {vec}(\mathbf{W})$. There is no prior knowledge of the channel sparsity rate, we thus combine EM-GAMP with channel sparsification to estimate the CSI adaptively after MC, and only 8\% RIS elements can receive signals. We will compare the proposed method with the following benchmarks:
\begin{itemize}
    \item OMP (orthogonal matching pursuit) adopted in \cite{11taha2019enabling} with prior on the channel sparsity rate and all RIS elements can receive signals. It is performed on only one subcarrier due to the high computational complexity.
    \item QIHT \cite{23jacques2013quantized} (quantized iterative hard thresholding), with prior on the channel sparsity rate, all RIS elements can receive signals.
    \item 1-bit CS \cite{68plan2012robust} (convex relaxation by sparse binomial regression), with prior on the channel sparsity rate, and all RIS elements can receive signals.
\end{itemize}

All the schemes above also adopt channel sparsification and 1-bit quantization for fair comparison. The pilots are OFDM symbols consisting of shifted ZC sequences. Since the length $N_p$ of pilots is limited in practice, we consider $N_p=16$ in Fig.~2 and $N_p=32$ in Fig.~3.

\begin{figure}[h]
  \centering
  \includegraphics[width=3.2in]{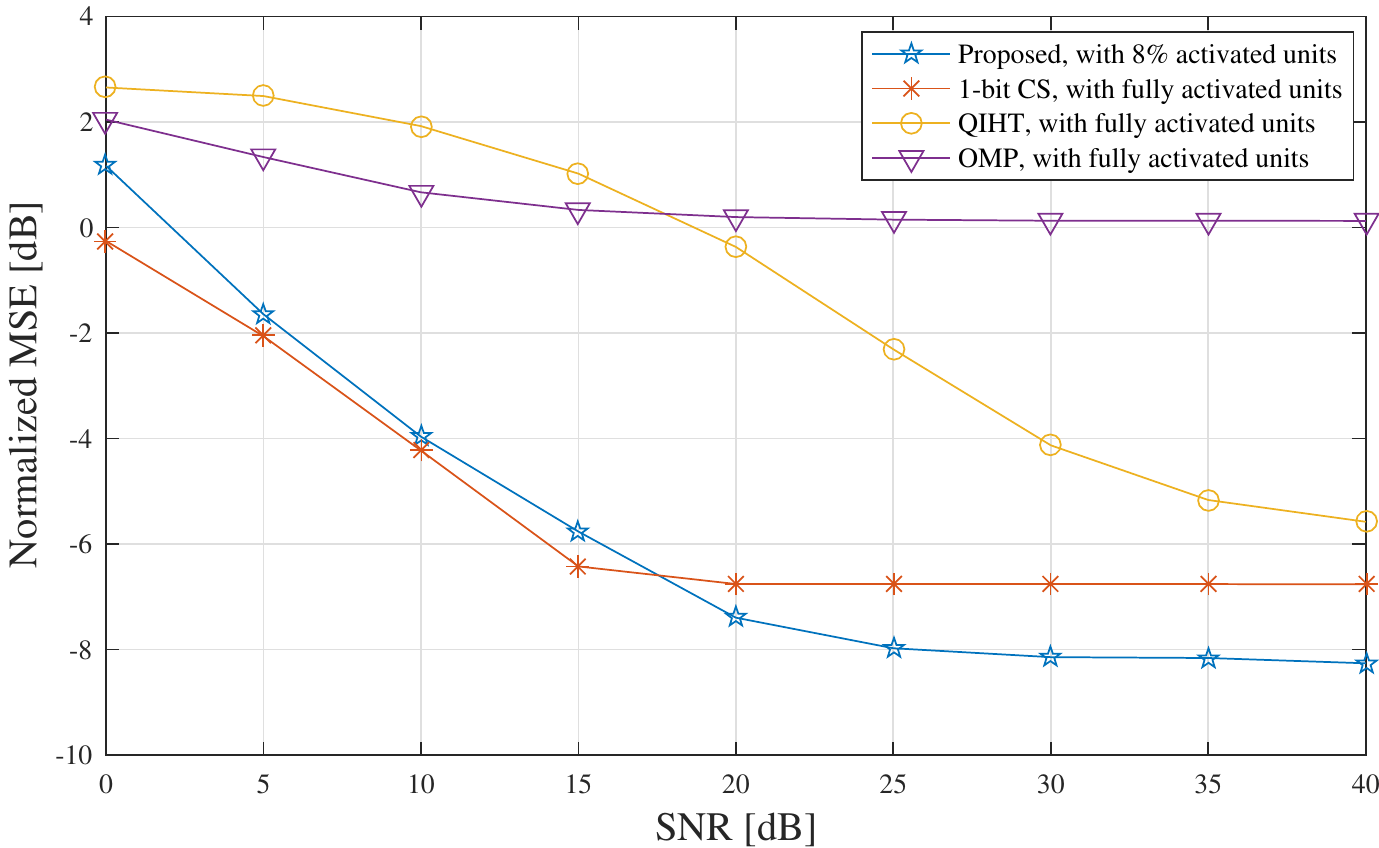}\\
  \caption{Channel estimation performance versus SNR, pilot length ${N_p} = 16$.} 
\end{figure}

\begin{figure}[h]
  \centering
  \includegraphics[width=3.2in]{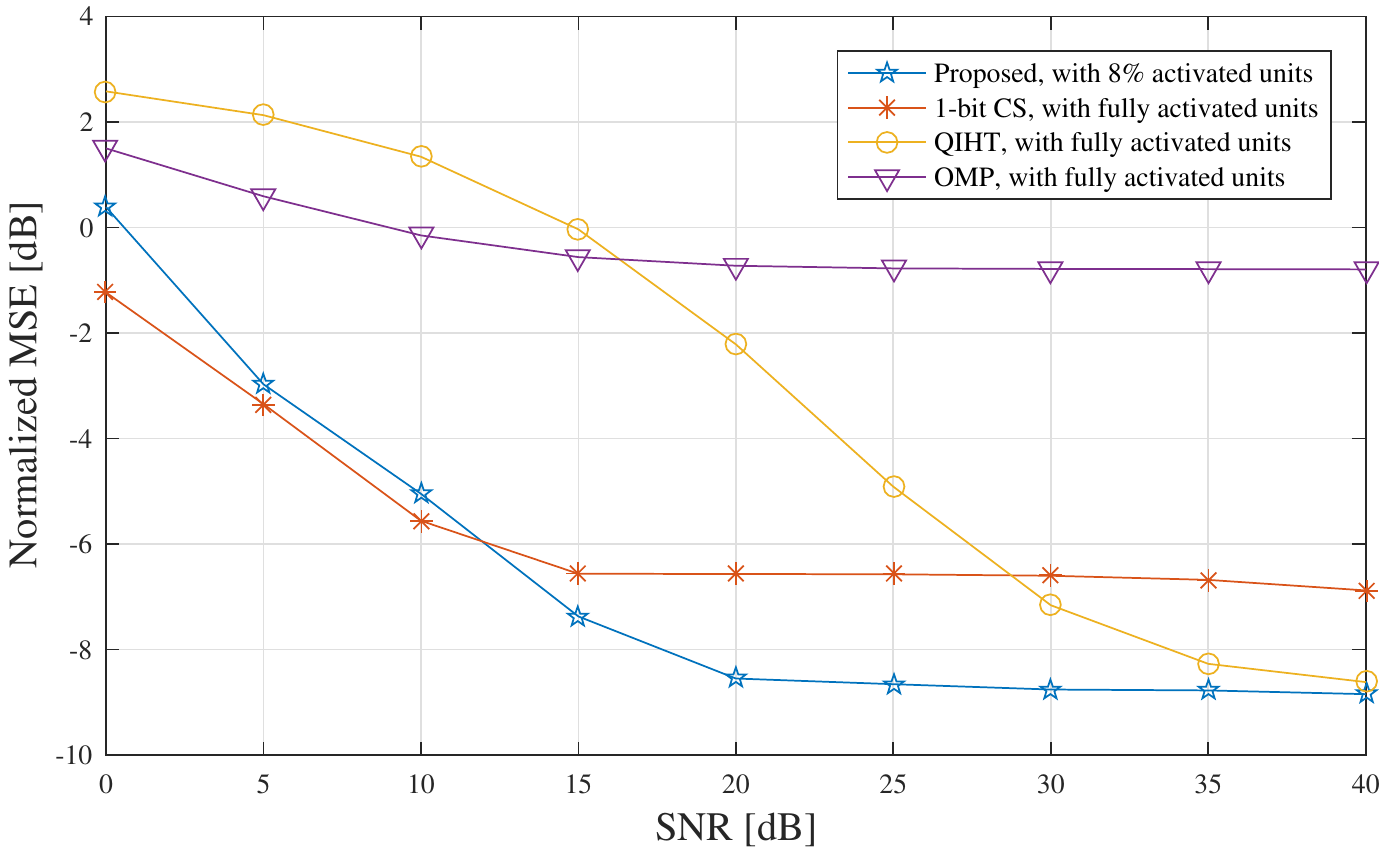}\\
  \caption{Channel estimation performance versus SNR, pilot length ${N_p} = 32$.} 
\end{figure}

We can observe in both Fig.~2 and Fig.~3 that our proposed scheme outperforms all the benchmarks when SNR in sufficiently large, even if only 8\% RF chains are assumed and without any prior on the channel sparsity rate. At lower SNRs, the performance of our scheme is very close to the 1-bit CS but  still better than OMP and QIHT. This is because the excessive noise affects the accuracy of MC, thus the EM process embedded in EM-GAMP cannot estimate the sparsity rate very well, which, together with the noise itself will impact the channel estimation accuracy of the GAMP algorithm. 

When $N_p$ increases from 16 to 32, we observe that there is only a limited performance improvement for the proposed scheme and the 1-bit CS algorithm, which indicates that the shorter pilot length may be sufficient. However, the performance of QIHT only improves in high SNR region, which is mainly because its hard thresholding is not adaptive to the increase of noise. The performance of OMP is relatively poor in both cases, which indicates its lack of robustness to quantization noise and that its pilot length may be not enough. 

\section{Conclusion}

In this paper, we have proposed a novel channel estimation scheme for a large-scale RIS-aided broadband mmWave MIMO communication system. By exploiting the joint sparsity of broadband mmWave channel in both the spatial and frequency domains, as well as its low-rankness property, we presented an efficient algorithm with fast implementation. In particular, we proposed an ADMM-based quantized MC approach, combined with EM-GAMP after a channel sparsification operation for the hardware/software joint design architecture. Simulation results show that our scheme outperforms the benchmarks in most cases, with less prior information and fewer hardware components. Equipped with only 8\% simplified receiver units, the proposed method can achieve more accurate channel estimation than the traditional methods where all RIS elements are active to receive signals.

\bibliographystyle{IEEEtran}
\bibliography{ref}
\end{document}